\documentclass[12pt,amsmath,amssymb,aps,nofootinbib]{revtex4-1}
\usepackage{graphicx,epsfig,dcolumn,bm,epic,eepic,float}
\usepackage{amsmath}
\usepackage[T1]{fontenc}
\usepackage[utf8]{inputenc}
\usepackage{mathtools} 
\usepackage{pifont}
\usepackage{amsmath}
\usepackage{latexsym}
\usepackage{color}
\usepackage{cancel}
\usepackage{scrextend}
\usepackage{hyperref}
\usepackage{amsmath}
\usepackage{amsthm}
\usepackage{eucal}
\usepackage{amssymb}
\usepackage{mathrsfs}
\usepackage[bottom]{footmisc}
\usepackage{makeidx,shortvrb,latexsym}
\usepackage{epstopdf}
\usepackage{mathtools}
\usepackage{ragged2e}
\usepackage[T1]{fontenc}
\newcommand {\be}{\begin{equation}}
\newcommand {\ee}{\end{equation}}
\newcommand {\ba}{\begin{eqnarray}}
\newcommand {\ea}{\end{eqnarray}}
\newcommand {\bea}{\begin{eqnarray}}
\newcommand {\eea}{\end{eqnarray}}

\pdfoutput=1

\pdfoutput=1

\begin{document}

\title{Signature of the Maximally Symmetric 2HDM via
 $W^{\pm}/Z$-Quadruplet Productions at the LHC}

\author{N. Darvishi$^{\ast,\dagger}$ and M.R. Masouminia$^{\ddagger}$}
\affiliation{\it \footnotesize
$^{\ast}$School of Physics and Astronomy, The University of Manchester, Manchester M13 9PL, United Kingdom
\\
$^{\dagger}$Faculty of Physics, University of Warsaw, Pasteura 5, 02-093 Warsaw, Poland
\\
$^{\ddagger}$Institute for Particle Physics Phenomenology, Department of Physics, Durham University, Durham DH1 3LE, United Kingdom
}

\begin{abstract}
${}$

\centerline{\bf ABSTRACT} \medskip

\noindent
We consider the Maximally Symmetric Two-Higgs Doublet Model (MS-2HDM) in which the so-called Standard Model (SM) alignment can be achieved naturally by the virtue of an SO(5) symmetry imposed on the 2HDM. We investigate the signature of the MS-2HDM via $p p \to HX \to VV^* X$ and $p p \to HHX \to VV^*V'V'^* X$ processes at the LHC for different values of $\tan \beta$. We perform our calculations with NLO QCD accuracy, using the \textsf{Herwig~7} multi-purpose event generator at $\sqrt{s}=13$~TeV center-of-mass energy. We show that the production of single SM-like Higgs bosons via $W^{\pm}/Z$-pairs is completely aligned with the SM. Interestingly, the presence of the heavy Higgs states significantly enhances the cross-section for the $W^{\pm}/Z$-quadruplet production channels in the low-$p_{\perp}$ regions. These vital analyses may aid the future discovery of this minimal and very predictive extension of the SM and can be generalised to other realisations of the 2HDM.

\end{abstract}

\maketitle

\section{Introduction}
\label{sec:intro}


For many years, the Standard Model (SM) has been the cornerstone for our understanding of the fundamental interactions of Particle Physics~\cite{Glashow:1961tr, Goldstone:1962es, Weinberg:1967tq, Salam:1968}. This was brought to its climax with the discovery of the Higgs boson at CERN's Large Hadron Collider (LHC)~\cite{Englert:1964et, Higgs:1964pj}. The data collected from this discovery imposes constraints over the coupling strengths of the Higgs boson, primarily to the electroweak (EW) gauge bosons ($V=W^{\pm}, \; Z$), which are very close to SM predictions~\cite{ATLAS:2019aqa, CMS:2019vxb}. Despite all these achievements, the SM still falls short of answering some of the most profound questions such as the origin of the observed matter-antimatter asymmetry and the dark matter relic abundance in the Universe. This has fueled numerous theoretical and experimental scrutinise in the study of theories Beyond the SM (BSM), particularly for models with extended Higgs sectors. These new-born theories regardless of their motivational and structural differences must restore those predictions of the SM that are consistent with the LHC observations. This is only possible within the so-called SM alignment limit~\cite{Ginzburg:1999fb,Chankowski:2000an,Delgado:2013zfa,Carena:2013ooa,Dev:2014yca,Bernon:2015qea,Darvishi:2016tni,Darvishi:2017bhf,Benakli:2018vqz,Lane:2018ycs}.


One of the simplest extensions of the SM is the Two-Higgs Doublet Model (2HDM), which enriches the SM scalar sector by introducing a second complex scalar doublet~\cite{Lee:1973iz, Pilaftsis:1999qt, Ginzburg:1999fb, Branco:2005em, Delgado:2013zfa}. This extension can provide new sources of CP violation~\cite{Lee:1973iz,Pilaftsis:1999qt}, introduce stable scalar DM candidates~\cite{Silveira:1985rk,Bonilla:2014xba,Krawczyk:2015xhl}, and give rise to EW baryogenesis~\cite{Kuzmin:1985mm, Cohen:1993nk}. Interestingly, the potential of this model contains the maximum number of $13$ distinct SU(2)$_L$-preserving accidental symmetries as subgroups of the maximal symmetry Sp(4)~$\sim$~SO(5)~\cite{Pilaftsis:2011ed, Battye:2011jj}. Thereby, the most minimal version of the 2HDM is an SO(5)-invariant potential, the so-called Maximally Symmetric 2HDM (MS-2HDM)~\cite{Pilaftsis:2011ed,Battye:2011jj,Dev:2014yca,Pilaftsis:2016erj,Dev:2017org,Hanson:2018uhf,Darvishi:2019ltl,Darvishi:2019dbh,Darvishi:2020teg,Birch-Sykes:2020btk}. In MS-2HDM, the SM alignment can emerge naturally as a consequence of an accidental SO(5) symmetry in the Higgs sector. This symmetry can be broken explicitly by the renormalization group (RG) effects and softly by the bilinear scalar mass term~$m^2_{12}$. A remarkable feature of this model is that all quartic couplings can unify at very large scales $\mu_X\sim 10^{9}\,$--$\,10^{20}$\, GeV, for a wide range of $\tan\beta$ values and charged Higgs-boson masses~\cite{Darvishi:2019ltl, Darvishi:2020teg}. This unique feature aids to gain a minimal and very predictive model which is governed only by three parameters: the quartic coupling unification scale $\mu_X$, the mass of charged Higgs $M_{h^{\pm}}$ (or $m^2_{12}$) and the value of $\tan\beta$. These three parameters allow one to determine the entire Higgs-mass spectrum of the model.  


The production of the EW gauge vector boson pairs and quadruplets have always been among the critical observations in the on-going attempts to probe the SM and search for signs of BSM physics at the LHC, within its high-energy hadronic scattering data. Also, these events have played a key role in the LHC precision measurements as well as the estimation of the irreducible backgrounds in Higgs boson searches. Moreover, since these processes are amongst the largest Higgs-tagged signatures at the current LHC energies, observing a distinct deviation from the SM theoretical predictions may be directly interpreted as a signal for BSM physics~\cite{Darvishi:2016fwo, Darvishi:2019uzp}. Therefore, it would be interesting to look for the possible signature of the MS-2HDM at the LHC via $W^{\pm}/Z$-pair and -quadruplet production events. In this paper, firstly we ensure that the predictions of the MS-2HDM for the $p p \to HX \to VV^* X$ production rates are aligned with their SM counterparts. Then, we evaluate the signature of the MS-2HDM via $p p \to HHX \to VV^*V'V'^* X$ events at the LHC for different values of $\tan \beta$. We calculate these production rates up to one QCD loop and 2 jets, using \textsf{Herwig 7} (v7.2.1) event generator \cite{Bahr:2008pv,Bellm:2015jjp,Bellm:2017bvx,Bellm:2019zci}. The corresponding amplitudes are provided by \textsf{MadGraph5} (v2.7.3) \cite{Alwall:2014hca} and matched to the NLO corrections using \textsf{Matchbox} \cite{Platzer:2011bc,Bellm:2019wrh}. The produced underlying events are showered by an AO \textit{MC@NLO} matched \textit{QCD+QED+EW}\footnote{The \textit{QCD+QED+EW} parton shower scheme is a new addition to \textsf{Herwig 7} which is introduced in~\cite{Masouminia:2021} and will be available to the public with the v7.3.0 release.} parton shower~\cite{Frixione:2002ik,Masouminia:2021}. Finally, the results of these simulations have been analyzed using \textsf{Rivet} (v3.1.1) \cite{Buckley:2010ar}.


The layout of the paper is as follows. After this introductory section, Section \ref{sec:alignment} briefly reviews the basic features of the 2HDM and the naturally aligned MS-2HDM. We also outline the Higgs-mass spectrum and our misalignment predictions for Higgs-boson couplings to gauge bosons. Section \ref{sec:Framework} shows the dominant channels for $p p \to HX \to VV^* X$ and $p p \to HHX \to VV^*V'V'^* X$ production events at the LHC and highlights the sup-processes where the new heavy Higgs bosons can substantially modify these cross-sections. This section also includes the calculation setup for our analysis. In Section \ref{sec:Results}, we discuss our numerical results for single and double Higgs production events in the MS-2HDM. Particularly, we show that the cross-section of the $p p \to HHX \to VV^*V'V'^* X$ processes is significantly enhanced with respect to the SM. Finally, Section \ref{sec:Conclusions} contains our conclusions.

\section{Type-II 2HDM and SM Alignment} \label{sec:alignment}

The 2HDM contains two scalar iso-doublets, $\Phi_{i}\,(i=1,2)$, with $\mathrm{U(1)_Y}$ hypercharges ${Y_{\Phi_{i}}=+1/2}$, as
\begin{equation}
{\Phi}_i\ =\ \left(
 \begin{matrix}
\Phi_i^+\\
 \Phi_i^0
 \end{matrix}\right)
\end{equation} 
The most general renormalisable 2HDM potential in terms of these doublets may be conveniently written as
\begin{eqnarray}
  \label{eq:V2HDM}
V &=& \mu_1^2 ( \Phi_1^{\dagger} \Phi_1) + \mu_2^2 ( \Phi_2^{\dagger} \Phi_2) 
 - \Big[ m_{12}^2 ( \Phi_1^{\dagger} \Phi_2)\: +\: {\rm H.c.}\Big] \nonumber \\
 &+& \lambda_1 ( \Phi_1^{\dagger} \Phi_1)^2 + \lambda_2 ( \Phi_2^{\dagger} \Phi_2)^2
 + \lambda_3 ( \Phi_1^{\dagger} \Phi_1)( \Phi_2^{\dagger} \Phi_2)
 + \lambda_4 ( \Phi_1^{\dagger} \Phi_2)( \Phi_2^{\dagger} \Phi_1) \nonumber \\
 &+& \bigg[\, {1 \over 2} \lambda_5 ( \Phi_1^{\dagger} \Phi_2)^2
 + \lambda_6 ( \Phi_1^{\dagger} \Phi_1)( \Phi_1^{\dagger} \Phi_2)
 + \lambda_7 ( \Phi_1^{\dagger} \Phi_2)( \Phi_2^{\dagger} \Phi_2)\: +\:
     {\rm H.c.} \bigg]\;,
\end{eqnarray}
where the mass terms $\mu^2_{1,2}$ and quartic couplings $\lambda_{1,2,3,4}$ are real parameters.  Instead, the remaining mass term $m^2_{12}$ and the quartic couplings $\lambda_{5,6,7}$ are complex. Out of these 14 theoretical parameters, only 11 are physical, since 3 parameters can be transformed away using a SU(2) reparametrisation of the Higgs doublets. Here, we restrict our attention to CP
conservation and to CP-conserving vacua.  In the Type-II 2HDM, both
scalar doublets receive nonzero vacuum expectation values (VEVs), as
$\langle \Phi_1 \rangle= (0,\, v_1/\sqrt{2})^\mathsf{T}$ and $\langle \Phi_2 \rangle= (0,\, v_2/\sqrt{2})^\mathsf{T}$.
Following the standard linear expansion of the scalar doublets about their VEVs, these can be re-expressed as
\begin{equation}
\Phi_i \  =\ \left(
 \begin{matrix}
 \phi_i^+ \\
 {1 \over \sqrt{2}} (v_i+ \phi_i + i\phi_i^0)
 \end{matrix}
 \right)\;.
\end{equation}
Accordingly, the minimization conditions resulting from the 2HDM
potential in~\eqref{eq:V2HDM} take on the following forms:
\begin{eqnarray}
\mu_{1}^2\ &=&\ m_{12}^2 {v_2\over v_1} - {1\over 2} v_1^2  \big(2
            \lambda_1+3 \lambda_6 {{v_2\over v_1} } + \lambda
            _{345}{({v_2\over v_1}) ^2}+\lambda_7 {({v_2\over v_1})^3} \big)\,,\\ 
\mu_{2}^2\ &=&\ m_{12}^2 {v_1\over v_2} - {1\over 2} v_1^2 \big(2
            \lambda_2+3 \lambda_6 {v_1\over v_2}  + \lambda
            _{345}({v_1\over v_2})^2+\lambda_7 ({v_1\over v_2})^3\big)\,, 
\end{eqnarray}
where ${\lambda }_{345}\equiv {\lambda }_3+{\lambda }_4+{\lambda }_5$, ${\lambda }_{5,6,7}\equiv {\rm Re}\,{\lambda }_{5,6,7} $ and ${m}_{12}^2\equiv {\rm Re}\,{m }_{12}^2 $. Also, the combination of $v_1$ and $v_2$ form the VEV of the SM doublet, $v^2 \equiv {(v_1^2 + v_2^2)}$ and their ratio is read $t_\beta\equiv\tan\beta =v_2/v_1$.  Therefore, the mixing  in the CP-even, CP-odd and charged scalar sectors can be individually governed by the same mixing angle $\beta$ as
\begin{equation}
\begin{pmatrix}
H_{\rm SM}\\ H_{p} 
\end{pmatrix}
=
\mathcal{O}(\beta)
\begin{pmatrix}
\phi_1 \\ \phi_2 
\end{pmatrix},
\quad 
\begin{pmatrix}
G^0 \\ a
\end{pmatrix}
=
\mathcal{O}(\beta)
\begin{pmatrix}
\phi_1^0 \\ \phi_2^0
\end{pmatrix},
\quad
\begin{pmatrix}
G^{\pm} \\ h^{\pm}
\end{pmatrix}
=
\mathcal{O}(\beta)
\begin{pmatrix}
\phi_1^{\pm} \\ \phi_2^{\pm}
\end{pmatrix},
\end{equation}
where the rotation matrix may be defined as below,
\begin{equation}
\mathcal{O}(\beta) = 
\begin{pmatrix}
c_\beta  & s_\beta
\\ 
-s_\beta &c_\beta
\end{pmatrix},
\end{equation}
with $c_\beta\equiv\cos \beta$ and $ s_\beta\equiv\sin\beta$.

After the spontaneous symmetry breaking, the standard $W^{\pm}$ and $Z$ bosons acquire their masses from the three would-be Goldstone bosons $(G^{\pm},G^0)$~\cite{Glashow:1961tr,Kita:2011yv}.  Thereafter, the model remains with five physical scalar states: two CP-even scalars ($H$ and $h$), one CP-odd scalar $(a)$ and two charged bosons ($h^{\pm}$). 
The masses of the $h^\pm= -s_\beta
             \phi_1^{\pm} + c_\beta \phi_2^{\pm}$ and $a=-s_\beta \phi_1^0 + c_\beta
         \phi_2^0$ scalars are given by
\begin{eqnarray}
M_{h^{\pm}}^2 \ &=&\ {m_{12}^2 \over s_{\beta} c_{\beta}} - {v^2 \over 2} (\lambda_4 + \lambda_5)
 + {v^2 \over 2 s_{\beta} c_{\beta}} (\lambda_6 c_{\beta}^2 
 + \lambda_7 s_{\beta}^2),
\nonumber \\
M_a^2 \ &=&\ M_{h^{\pm}}^2 + {v^2 \over 2} (\lambda_4 - \lambda_5),
\end{eqnarray}
 Moreover, the masses of the physical states $H$ and $h$ may be obtained by diagonalising the CP-even $\phi_1$ and $\phi_2$ mass matrix~$M^2_S$,
\begin{equation}
M_S^2=\left(
\begin{matrix}
A\,& C\\
C\,& B
 \end{matrix}\right),
 \label{abc}
\end{equation}
which may be explicitly written in the following form
 \begin{align}
 A=& M_a^2  s_{\beta}^2 + v^2 \left(  2 \lambda_1 c_{\beta}^2 + \lambda_5 s_{\beta}^2 + 2 \lambda_6 s_{\beta} c_{\beta}  \right),
 \nonumber \\ 
 B=& M_a^2 c_{\beta}^2+ v^2 \left( 2 \lambda_2 s_{\beta}^2 + \lambda_5 c_{\beta}^2 + 2 \lambda_7 s_{\beta} c_{\beta}\right),
 \nonumber \\ 
 C=& -M_a^2 s_{\beta}c_{\beta} +v^2 \left(  (\lambda_{3} + \lambda_{4}) s_{\beta} c_{\beta} + \lambda_6 c_{\beta}^2 + \lambda_7 s_{\beta}^2 \right). \nonumber
 \end{align}
The physical states $h$ and $H$ can be given in terms of the mixing angles $\alpha$, as
\begin{equation}
\begin{pmatrix}
H \\ h
\end{pmatrix}
=
\mathcal{O}(\alpha)
\begin{pmatrix}
\phi_1 \\ \phi_2 
\end{pmatrix}=
\mathcal{O}(\alpha)\mathcal{O}(\beta)^{-1}
\begin{pmatrix}
H_{\rm SM}\\ H_p 
\end{pmatrix}.
\end{equation}
 Hence, the SM Higgs field may now be identified by the following linear fields combination,
\begin{equation}
   \label{g-c}
H_{\text{SM}}\ =\ \phi_1 \cos \beta + \phi_2 \sin \beta\ 
 =\ H \cos (\beta - \alpha) + h \sin (\beta - \alpha)\; .
\end{equation}
Henceforth, the SM-normalised couplings of the CP-even $H$ and $h$ scalars to the EW gauge bosons are given by
\begin{align}
 g_{hVV} = \sin (\beta - \alpha)\;,
\qquad 
g_{HVV} = \cos (\beta - \alpha).
\end{align}

Thereby, there are two ways to realise exact SM alignment limit; (i) SM-like $h$ scenario with $\sin(\beta-\alpha) = 1$ and (ii) SM-like $H$ scenario with $\cos(\beta-\alpha) = 1$.
Here, we consider the second realization with $\beta=\alpha$ and assuming the CP-even scalar partner $h$ can be either lighter or heavier than the observed scalar resonance at the LHC. As it can be seen in  \eqref{g-c}, with this assumption, the SM-like Higgs boson becomes aligned
with one of the neutral eigenstates.
So, the CP-even mass matrix $M_S^2$ takes on the form
 \begin{eqnarray}
\widehat{M}_S^2 & = & \mathcal{O}(\beta) M_S^2 \mathcal{O}(\beta)^{-1}= \left(
 \begin{matrix}
 \widehat{A} & \widehat{C} \\
 \widehat{C} & \widehat{B}
 \end{matrix}
 \right),
\end{eqnarray}
with
\begin{eqnarray}
\widehat{A} &=& 2v^2 \left[ c_{\beta}^4 \lambda_1 + s_{\beta}^2 c_{\beta}^2 \lambda_{345}
 + s_{\beta}^4 \lambda_2 + 2 s_{\beta} c_{\beta} \left( c_{\beta}^2 \lambda_6 
 + s_{\beta}^2 \lambda_7 \right) \right], 
\nonumber \\
\widehat{B} &=& M_a^2 + \lambda_5 v^2 + 2v^2 \left[ s_{\beta}^2 c_{\beta}^2 \left(
 \lambda_1 + \lambda_2 - \lambda_{345} \right) - s_{\beta} c_{\beta} \left(
 c_{\beta}^2 - s_{\beta}^2 \right) \left(\lambda_6 - \lambda_7 \right) \right],
 \\ 
\widehat{C} &=& v^2 \left[ s_{\beta}^3 c_{\beta} \left( 2 \lambda_2 - \lambda_{345} \right)
 - c_{\beta}^3 s_{\beta} \left( 2 \lambda_1 - \lambda_{345} \right)
 + c_{\beta}^2 \left( 1 - 4 s_{\beta}^2 \right) \lambda_6
 + s_{\beta}^2 \left(4 c_{\beta}^2 - 1 \right) \lambda_7 \right].
\nonumber 
\end{eqnarray}
From the above relations, we may observe that the SM alignment limit, $\beta=\alpha$, can be understood if either $\widehat{C} \to 0$ or (ii)~$M_{h^\pm}\!\sim\!M_a \gg v$. However, these can be obtained naturally  by imposing the symmetries of model to constraint the certain parameters required for the SM alignment limit. The 2HDM potential contains 13 accidental symmetries, which have been fully classified in ~\cite{Pilaftsis:2011ed,Darvishi:2019dbh,Birch-Sykes:2020btk}. Of these, eight restrict the quartic couplings such that the alignment condition $\widehat{C} \to 0$ is met. However, only for three symmetries exact alignment can be achieved naturally without imposing any constraints on the values of $\tan\beta$, nor on the bilinear mass terms $\mu^{2}_{1,2}$ and $m^{2}_{12}$~\cite{Pilaftsis:2016erj,Darvishi:2020teg}. In the simplest scenario, dubbed the MS-2HDM, the SM alignment can be naturally realised as a consequence of an accidental SO(5) symmetry in the Higgs sector~\cite{Pilaftsis:2011ed,Dev:2014yca,Dev:2017org,Darvishi:2019ltl}. In the MS-2HDM, the SO(5) symmetry puts severe restrictions on the allowed form of the kinematic parameters of the 2HDM potential in~\eqref{eq:V2HDM},
\begin{align}
\label{m-c}
\mu_1^2 = \mu_2^2\,, \quad m_{12}^2 = 0\,, \quad
\lambda_2 = \lambda_1\,, \quad\ \lambda_3 = 2\lambda_1\,, \quad 
\lambda_4 = \lambda_5 = \lambda_6 = \lambda_7 = 0\;.
\end{align}
These parameters produce one massive CP-even scalar with $M_H^2 = 2 \lambda_2 v^2$. 
The other four physical scalars $(h, a, h^{\pm})$ become massless pseudo-Goldstone bosons, 
that would participate in several Higgs decay channels which are inconsistent with observation. 
To avoid these, we consider that only the soft SO(5)-breaking parameter Re\,$m^2_{12}$ is 
non-zero~\cite{Dev:2014yca,Dev:2017org,Darvishi:2019ltl}.
With this minimal addition to the MS-2HDM potential, the scalar-boson masses to a 
very good approximation are given by 
\begin{equation}
    \label{eq:m12}
 M_H^2 = 2\lambda_2 v^2, \qquad M_h^2\ =\ M_a^2\ =\ M_{h^{\pm}}^2\ 
 =\ {\,m_{12}^2 \over s_{\beta} c_{\beta}}\;.
\end{equation}
In addition to softly broken the SO(5) symmetry by the bilinear scalar mass term $m^2_{12}$, 
this symmetry can be broken explicitly by RG effects due to non-zero values of gauge coupling $g'$
and Yukawa couplings. The SO(5) symmetry may be realised at large RG scales $\mu_X$ in two discrete values 
$\mu^{(1)}_X \sim 10^{11}$\,GeV and $\mu^{(2)}_X \sim 10^{18}$\,GeV by considering the 
threshold scales $\mu_{thr} = M_h^{\pm}=500$\,GeV. Nonetheless, assuming larger 
values for the threshold scale e.g. $\mu_{thr} = M_h^{\pm}= $1\,TeV, 10\,TeV and 100\,TeV the RG scales 
$\mu_X$ can reach to $\sim\,10^{20}$\,GeV~\cite{Darvishi:2019ltl}.
In this context, when this symmetry breaks due to RG effect, giving rise to calculable non-zero values 
for misalignment predictions of all Higgs boson couplings to SM particles.

In the EW scale, the $H$-boson couplings in terms of the light-to-heavy scalar-mixing 
parameter may be expressed by $\theta_\mathcal{S}\equiv\widehat{C}/\widehat{B}$.  
So, the approximate analytic expressions may be given by~\cite{Darvishi:2019ltl}
\begin{subequations} \label{ex-gc}
\begin{align}
g_{HVV}&\simeq 1-{\theta_\mathcal{S}^2\over 2},
\\
g_{hVV}&\simeq -{\theta_\mathcal{S}}\ =\ 
{v^2 s_{\beta} c_{\beta} \over M_a^2 +  v^2 \lambda_5 }\,
 \Big[\, c_{\beta}^2 \left( 2 \lambda_1 - \lambda_{345} \right) -
   s_{\beta}^2 \left( 2 \lambda_2 - \lambda_{345} \right)
   \Big]. 
\end{align}
\end{subequations}
Given the narrow experimental limits on the deviation of $g_{HVV}$ from~1, one must have the parameter
$\theta_\mathcal{S} \ll 1$.
In the SM alignment limit, we have $g_{Huu}\,\to\,1$ and
$g_{Hdd} \to$ 1. To this extent, an approximate analytic expressions for the 
$h$- and $H$-boson couplings to up- and down-type quarks  are given by
\begin{align}
g_{Huu}&\simeq  1+{t^{-1}_\beta}\,{\theta_\mathcal{S}}, \qquad \qquad
 g_{Hdd} \simeq 1-{\theta_\mathcal{S}}\,{t_\beta}, \\ \nonumber
 g_{huu}&\simeq -{\theta_\mathcal{S}}+{t^{-1}_\beta}, \qquad \qquad
 g_{hdd} \simeq -{\theta_\mathcal{S}}-{t_\beta} .
   \label{mfc}
\end{align}
Also, the trilinear Higgs couplings can be defined via the light-to-heavy scalar-mixing parameter $\theta_\mathcal{S}$, in the following forms
\begin{align}
\kappa_{HHH}&\simeq\ {3\over 2 v}\Big[\widehat{A}+({3\over 2 v}\widehat{A}- {\,m_{12}^2 \over s_{\beta} c_{\beta}})\theta_\mathcal{S}^2\Big],
\\
\kappa_{hHH}&\simeq\ 
{\theta_\mathcal{S} \over 2 v}\Big[-2\widehat{A}-\widehat{B}+2{\,m_{12}^2 \over s_{\beta} c_{\beta}}\Big],
\label{hhh}
\end{align}
where the $\kappa_{hHH}$ coupling vanishes in the exact alignment limit and $\kappa_{HHH}$ converges to Its SM counterpart.

\begin{table*}[t]
 \centering
 \small
  \begin{tabular}{c c c c c c c c c c c c c c c c c c c c c c c }
 \hline\hline
Couplings && ATLAS && CMS && $\tan\beta = 2$ && $\tan\beta = 5$ && $\tan\beta = 50$ \\ \hline
 $|g_{HZZ}^{\text{low-scale}}|$  && [0.86, 1.00]&& [0.90, 1.00] && 0.9999 && 0.9999 && 0.9999 \\
 $|g_{HZZ}^{\text{high-scale}}|$ && &&                          && 0.9981 && 0.9998 && 0.9999 \\ \hline
 $|g_{Htt}^{\text{low-scale}}|$  && $1.31^{+0.35}_{-0.33}$ && $1.45^{+0.42}_{-0.32}$ && 1.0049 && 1.0014 && 1.0000 \\
 $|g_{Htt}^{\text{high-scale}}|$ && &&                                               && 1.0987 && 1.0179 && 1.0001 \\ \hline
 $|g_{Hbb}^{\text{low-scale}}|$ && $0.49^{+0.26}_{-0.19}$ && $0.57^{+0.16}_{-0.16}$  && 0.9803 && 0.9649 && 0.9590 \\
 $|g_{Hbb}^{\text{high-scale}}|$&& &&                                                && 0.8810 && 0.9264 && 0.9427  \\ \hline
 $|g_{HHH}^{\text{low-scale}}|$  && [-5.1, 12.0]&& [-11.8, 18.8] && 0.9968 && 0.9968 && 0.9968 \\
 $|g_{HHH}^{\text{high-scale}}|$ &&                 &&                          && 0.9394 && 0.9936 && 0.9968 \\ \hline
\end{tabular}
\caption{\it Predicted values of the SM-like Higgs boson  couplings to the $Z$ boson and to top- and bottom-quarks in the MS-2HDM for both scenarios with low- and high-scale quartic coupling unification,
assuming $M_{h^{\pm}}=500$\,GeV. The corresponding central values for these couplings from ATLAS and CMS are also given, including their uncertainties \cite{Khachatryan:2016vau}. The ratio of the Higgs boson
self-coupling to its SM value, $g_{HHH}$,  is constrained at 95$\%$ confidence level in
observation (expectation) to $-5.0<g_{HHH} <12.0 \, (-5.8 <g_{HHH}<12.0)$ and $-11.8<g_{HHH}<18.8$ 
$(-7.1 <g_{HHH}< 13.6)$ by ATLAS and CMS, respectively~\cite{Aad:2019uzh,Sirunyan:2018ayu}. }
\label{ex}
\end{table*}

The MS-2HDM is a minimal and very predictive extension of the SM governed by only three additional parameters: the unification scale~$\mu_X$, the charged Higgs mass $M_{h^{\pm}}$ and $\tan\beta$, allowing one to determine the entire Higgs sector of the model. 
Previously, we presented our MS-2HDM benchmarks in terms of these input parameters, for our misalignment predictions of the SM-like Higgs-boson couplings to the $W^\pm$ and $Z$ bosons~\cite{Darvishi:2019ltl}. These benchmarks are given in Table \ref{ex}. 
Having considered the running mass $M_{H}(M_{h^{\pm}}=500)$ to obtain $M_{H}(m_t)\sim 125$ GeV, the mass of heavy neutral Higgses are $M_{h}\sim498$ GeV and  $M_{a}\sim 495$ GeV.

Here, we intend to calculate the rate of $W^{\pm}/Z$-pair and -quadruplet productions through single and double SM-like Higgs boson decay modes based on these benchmarks. 

\section{$W^{\pm}/Z$-Quadruplet  Production and Even generation Framework}
\label{sec:Framework}

The $W^{\pm}/Z$-pair and -quadruplet production events are amongst the largest Higgs-tagged signatures at the current LHC energies. However,  the best processes for probing the sings of an extended Higgs sector are the $W^{\pm}/Z$- quadruplet productions through double Higgs decays, i.e. $p p \to HH \to VV^*V'V'^*$. This is since, as we have shown in Section~\ref{sec:alignment},  the SM-like Higgs boson in the MS-2HDM couples to the EW gauge bosons with coupling strength exactly as that of the SM Higgs boson, while the other neutral heavy states do not couple to them at all~\cite{Darvishi:2019ltl}. On the other hand, because of the mixing between the light and the heavy states, the $W^{\pm}/Z$- quadruplet production through double Higgs decays may be enhanced. Henceforth, we calculate these processes within the MS-2HDM, which can include Higgs trilinear decay modes $H/h\to HH$.
The prediction for the relative trilinear Higgs self-coupling to its SM value, $g_{HHH}$, for different values of $\tan\beta$ which are allowed by the observed (expected) data are summarized in Table \ref{ex}. The $g_{HHH}$  is constrained at 95$\%$ confidence level in
observation (expectation) to $-5.0<g_{HHH}<12.0$ $(-5.8 <g_{HHH}< 12.0)$ and $-11.8<g_{HHH}<18.8 \,
(-7.1<g_{HHH} <13.6)$ by ATLAS and CMS, respectively~\cite{Aad:2019uzh,Sirunyan:2018ayu}. 
Additionally, the non-zero values for misalignment predictions giving rise to the sensitive BSM decay mode $h\to HH$. The observed upper limit on the resonant production cross-section times the branching fraction of $h \to HH$ with ATLAS detector ranges between $40$\,pb and 6.1 pb, while the expected limit ranges between 17.6 pb and 4.4 pb, for a hypothetical resonance with a mass in the range of 260-500 GeV~\cite{Aaboud:2018ewm}. Our prediction for the resonant production $\sigma(gg \to h) \times Br(h \to HH)$ for $\tan\beta=2,\,5$ and $50$ are ranged between 4.3 pb and 4.0 pb. 

Note that there are several interesting channels that can be considered when looking for the signal of the MS-2HDM model at the LHC, namely single and double Higgs production events through $H \to b\bar{b}$, $H \to W^+W^-$, $H \to ZZ$ and $H \to \tau\bar{\tau}$ decay channels. Indeed, the $H \to b\bar{b}$ decay has the largest branching ratio of $\sim\%57$, but they are generally difficult to observe due to their large backgrounds. On the other hand, the $H \to W^+W^-$ and $H \to ZZ$ decays have the combined branching ratio of $\sim\%$25 and are very likely to produce clean experimental signatures via their leptonic decay channels $W^+W^- \to l^+\nu_l + l^{\prime -} \nu_{l'}$ and $ZZ \to l^+  l^- + l^{\prime +} l^{\prime -}$. 

\begin{figure}[t]
\centering
\includegraphics[width=.8\textwidth]{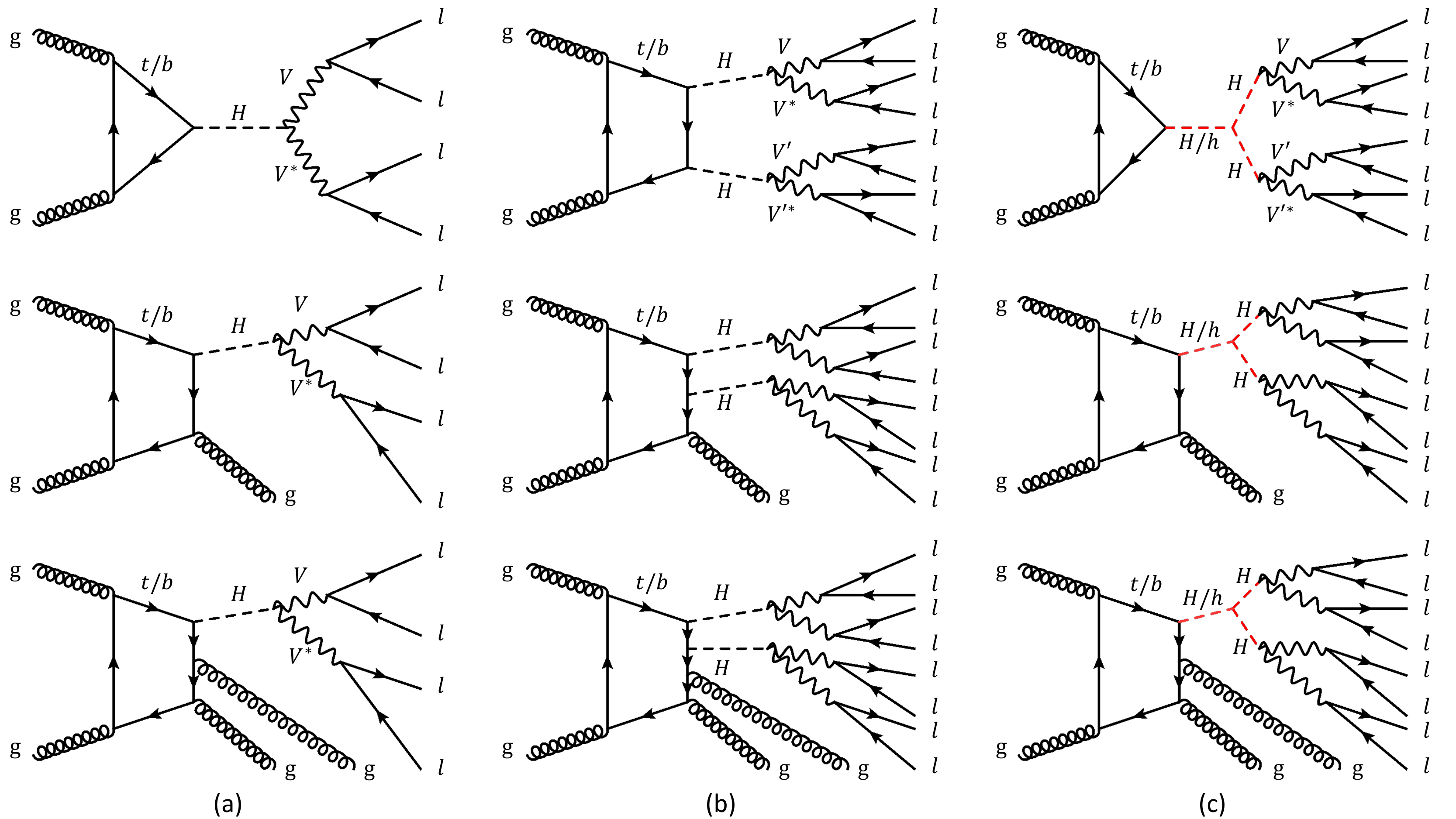}
\caption{ \it Dominant sub-processes for $W^{\pm}/Z$-pair and -quadruplet productions through single and double Higgs boson decays at the LHC up to 2 jets; (a)  $W^{\pm}/Z$-pair productions via single Higgs decay channels. (b)  $W^{\pm}/Z$-quadruplet productions through double Higgs boson decays. (c) Double Higgs production channels via $H/h\to H H$ decays.}
\label{fig1}
\end{figure}

Before delving into the details of this section, we check the $W^{\pm}/Z$-pair production events through single Higgs decays to ensure the MS-2HDM predictions are aligned with those of the SM.  The exclusive production of the EW gauge boson pairs and quadruplets via single and double Higgs decays can be given as

\begin{align}
p(p_1) + p(p_2) & \to  H(p_H) + X \nonumber \\
					& \to  W^+(p_{W^+}) + W^{-}(p_{W^-}) + X 
					  \to   l_1^+\nu_{l_1} + l_2^{-} \nu_{l_2} + X, 
\label{WW}
\\[0.1in]
p(p_1) + p(p_2) & \to  H(p_H) + X \nonumber \\
					& \to  Z(p_{1,Z}) + Z(p_{2,Z}) + X 
					  \to   l_1^+  l_1^- + l_2^{+} l_2^{-} + X,
\label{ZZ}
\\[0.1in]
p(p_1) + p(p_2) & \to  H(p_{1,H}) + H(p_{2,H}) + X  \nonumber \\
					& \to  W^+(p_{1,W^+}) + W^-(p_{1,W^-}) 
					+ W^+(p_{2,W^+}) + W^-(p_{2,W^-}) + X, \nonumber \\
					& \to  l_1^+\nu_{l_1} + l_2^{-} \nu_{l_2} 
					+ l_3^+\nu_{l_3} + l_4^{-} \nu_{l_4} + X, 
\label{WWWW}
\\[0.1in]
p(p_1) + p(p_2) & \to  H(p_{1,H}) + H(p_{2,H}) + X  \nonumber \\
					& \to  Z(p_{1,Z}) + Z(p_{2,Z}) 
					+ Z(p_{3,Z}) + Z(p_{4,Z}) + X, \nonumber \\
					& \to  l_1^+ l_1^- + l_2^+ l_2^- 
					+ l_3^+ l_3^- + l_4^+ l_4^- + X. 
\label{ZZZZ}
\end{align}
In the above processes, the Higgs boson production mechanism is dominated by the gluon fusion channels $gg \to H$ and $gg \to HH$, which account for $\sim$\%95 of the Higgs production rate at the LHC\footnote{In~\cite{Masouminia:2018dwj}, it has been shown that one can simply enhance the LO differential cross-sections of Higgs production with the use of a factorization-scale-dependant K-factor and forget about the higher-order and virtual corrections to these processes, e.g. via the $W/Z$ Higgs-strahlung sub-processes~\cite{Baglio:2010ae}. Nevertheless, for the sake of completeness, we are considering the full range of real and virtual QCD and QED contributions to \eqref{WW}, \eqref{ZZ}, \eqref{WWWW} and \eqref{ZZZZ}, using the \textsf{Matchbox} merging functionality within \textsf{Herwig 7}.}~\cite{Darvishi:2016fwo,Masouminia:2018dwj}. Figure~\ref{fig1} displays all dominant sub-processes up to two jets that are mediated through the exchange of a heavy virtual top/bottom quark. The parts (a) and (b) of Figure~\ref{fig1} show the dominant single and double Higgs production channels, respectively. The part (c) showcases the contributions of the light and heavy Higgs bosons through the trilinear Higgs vertices, where the signature of new physics would rise. Also, the leptonic decay channels $W^+W^- \to l^+\nu_l + l^{\prime -} \nu_{l'}$ and $ZZ \to l^+  l^- + l^{\prime +} l^{\prime -}$ have been considered to ensure a clean observable signature and to prevent the reconstruction of $H \to VV^*$ resonances. 

In our calculations for the production rates of the $p p \to HX \to VV^*$ and $p p \to H HX \to VV^*V'V'^*$ events, we utilize the \textsf{Herwig 7} (v7.2.1) event generator \cite{Bahr:2008pv,Bellm:2015jjp,Bellm:2017bvx,Bellm:2019zci}. This will be done for both the SM and the MS-2HDM, in the collinear factorization framework. The contributing matrix elements are generated with \textsf{MadGraph5} and convoluted by \textsf{MMHT2014} parton distribution function libraries~\cite{Harland-Lang:2014zoa} via \textsf{LHAPDF} interface~\cite{Buckley:2014ana}. The underlying events are enhanced by an AO \textit{QCD+QED+EW} parton shower scheme~\cite{Platzer:2009jq,Platzer:2011bc,Masouminia:2021} and the cluster model hadronization~\cite{Webber:1983if}. The generated events are then analysed with \textsf{Rivet}, based on the existing analysis \textsf{MC\_WWINC} and \textsf{MC\_ZZINC} which are modified according to our needs. 

\section{Numerical results and discussion}
\label{sec:Results}

In this section, we present our results for $W^{\pm}/Z$-pair and quadruplet production through the single and the double Higgs bosons decays at~$\sqrt{s}=13$~TeV. In the first step, we check the efficiency of our calculation setup by evaluating the single Higgs production results in the SM with the existing experimental observations from the CMS collaboration, including their
statistical and systematic uncertainties~\cite{Darvishi:2019uzp,Sirunyan:2020tzo,Sirunyan:2018sgc}.  

\begin{figure}[t]
\centering
\includegraphics[width=1\textwidth]{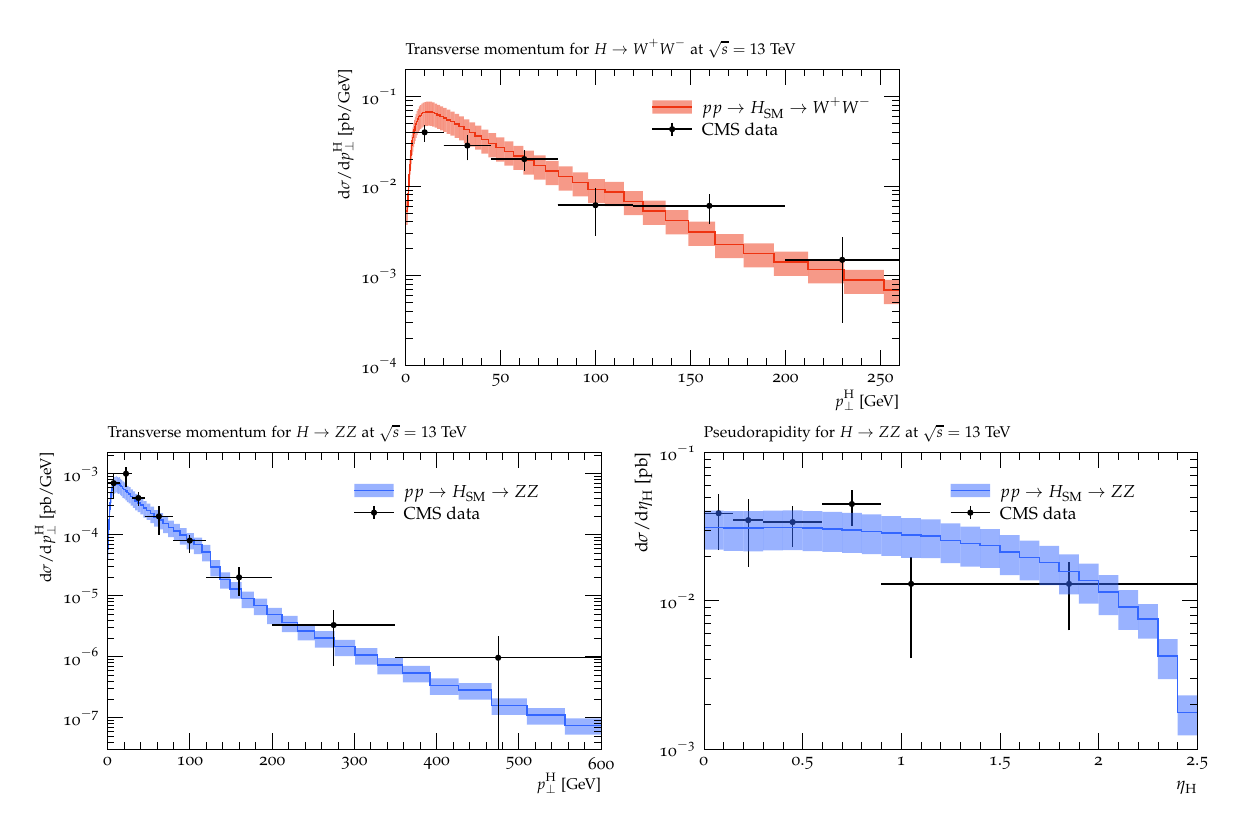}
\caption{\it Differential cross-section for single Higgs boson production as a function of its transverse momentum and pseudo-rapidity. The calculations are in the colliner framewok, using \textsf{Herwig 7} (v7.2.1) at $\sqrt{s}=13$~TeV. The top panel corresponds to the $pp \to H X\to W^+ W^- X$ channel while the bottom panels are for the $pp \to H X\to ZZ X$ channel. The data are from the CMS collaboration~\cite{Sirunyan:2020tzo,Sirunyan:2018sgc}. To calculate the uncertainty region, we have manipulated the factorization hard-scale by a factor of 2.}
\label{CMS13}
\end{figure}

\begin{figure}[t]
\centering
\includegraphics[width=1\textwidth]{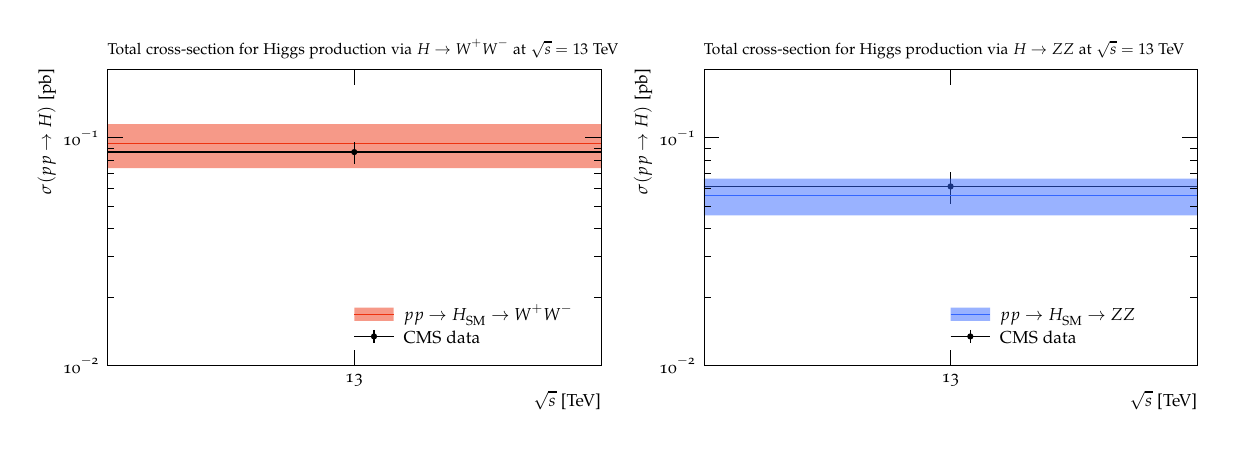}
\caption{\it Feducial cross-section for $W^{\pm}(Z)$-pair productions through single SM Higgs boson decays at the LHC for $\sqrt{s}=13$~TeV displayed in left (right) panel. The data are from the CMS collaboration, including their
statistical and systematic uncertainties~\cite{Sirunyan:2020tzo,Sirunyan:2018sgc}. To calculate the uncertainty region, we have manipulated the factorization hard-scale by a factor of 2.}
\label{FCD_CMS13}
\end{figure}

In Figure \ref{CMS13}, we exhibit the results of our analysis for single Higgs bosons production via $H\to VV^*$ decay channels. The top panel demonstrates the kinematically reconstructed transverse momentum distribution of the exchanged Higgs boson from the $H \to W^+W^-$ decay mode while the bottom panels correspond to the transverse momentum and pseudo-rapidity from the $H \to ZZ$ decay mode. Figure~\ref{FCD_CMS13} shows the fiducial cross-sections of single Higgs production through $W^{\pm}$-pairs (left panel) and $Z$-pairs (right panel). The event selection criteria for these calculations have been chosen in accordance with the reported conditions in~\cite{Sirunyan:2020tzo,Sirunyan:2018sgc}.
Despite the low precision of the experimental data, in both figures one can readily observe that these predictions are perfectly capable of describing the experimental measurements.   

\begin{figure}[t]
\centering
\includegraphics[width=1\textwidth]{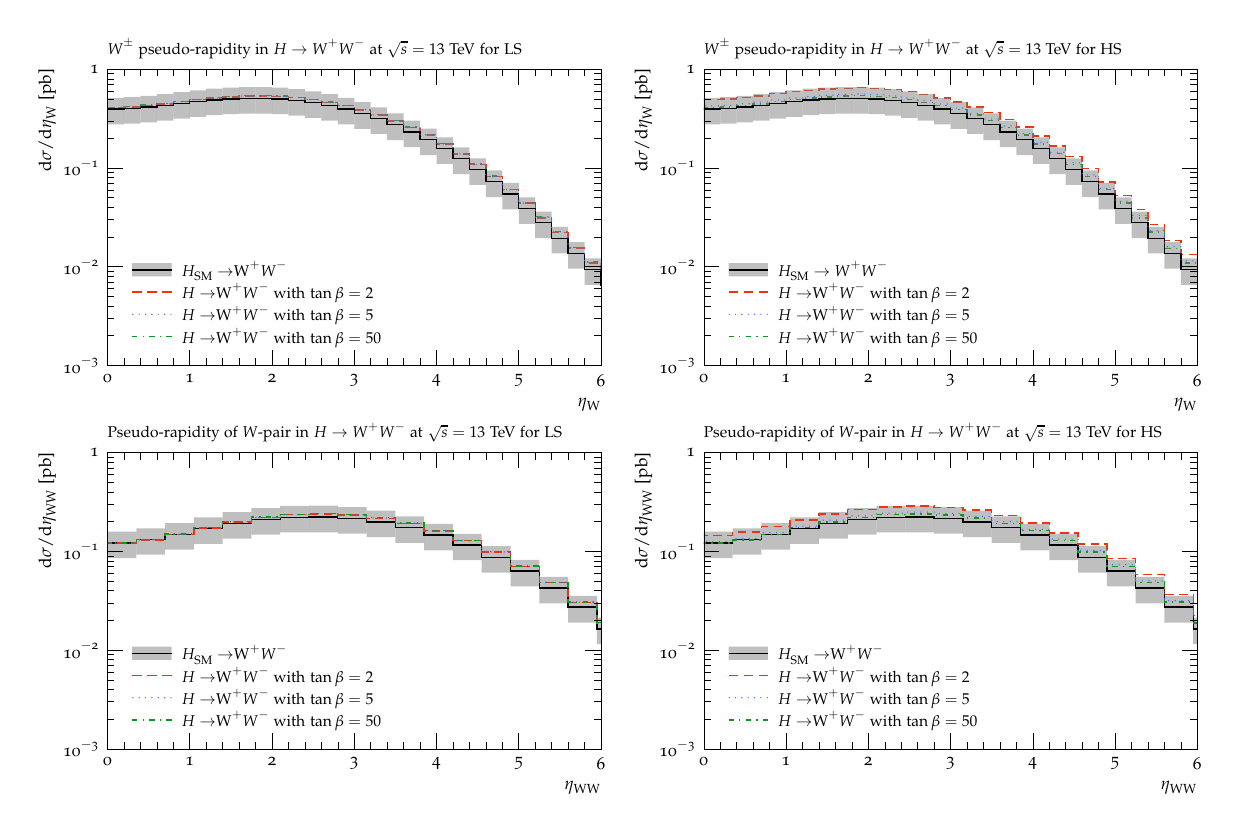}
\caption{\it Differential cross-section as a function of pseudo-rapidity for $W^{\pm}$-pair productions through single SM-like Higgs boson ($H$) events  displayed for lower-scale  (higher-scale) quartic coupling unification point in left panels (right panels).  These are shown for  different values of $\tan \beta$ at $\sqrt{s}=13$~TeV. The top panels correspond to the kinematic properties of the $W^{\pm}$ bosons while the bottom panels depict the behaviour of $W^{\pm}$-pairs. Note that, these results are compared with relevant theoretical predictions in the SM within its uncertainty bounds.}
\label{WW-y}
\end{figure}

\begin{figure}[t]
\centering
\includegraphics[width=1\textwidth]{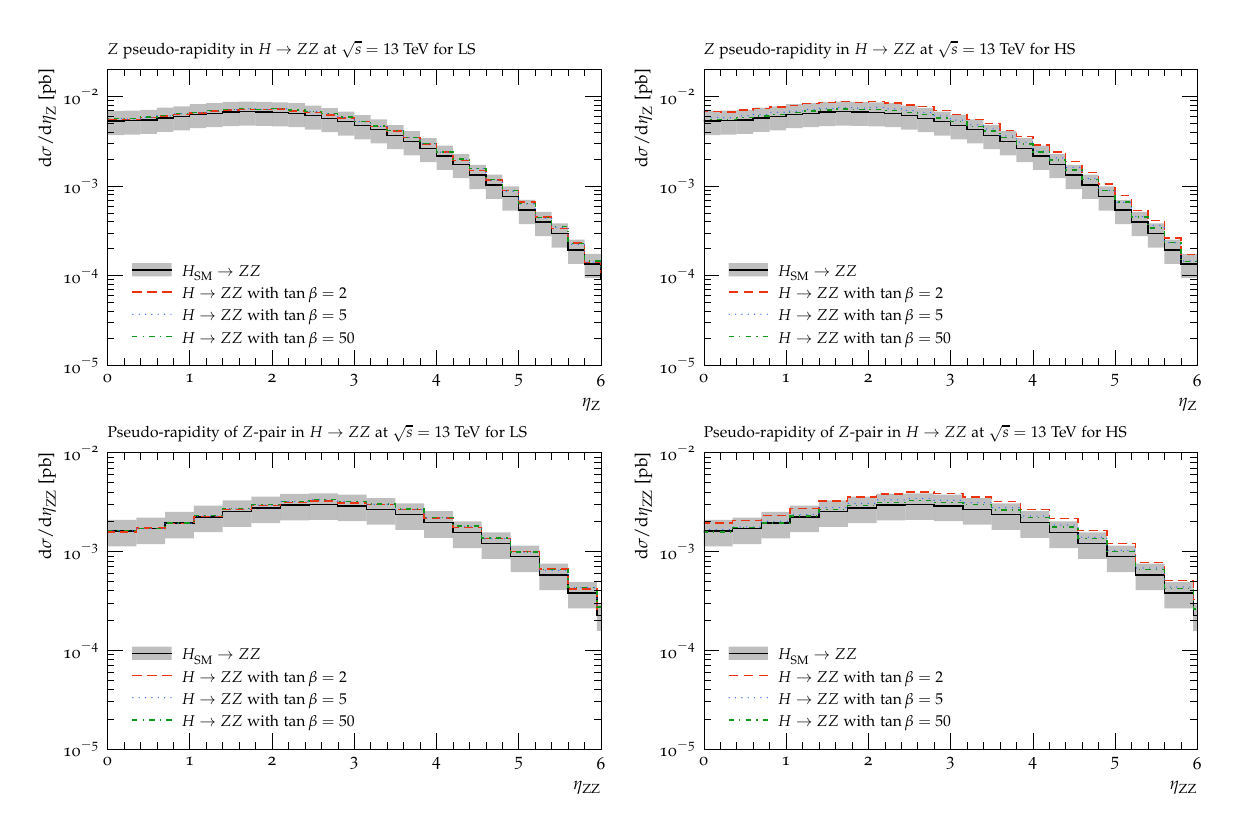}
\caption{\it The same as in Figure~\ref{WW-y}, but for $Z$-pair productions}.
\label{ZZ-y}
\end{figure}

\begin{figure}[t]
\centering
\includegraphics[width=1\textwidth]{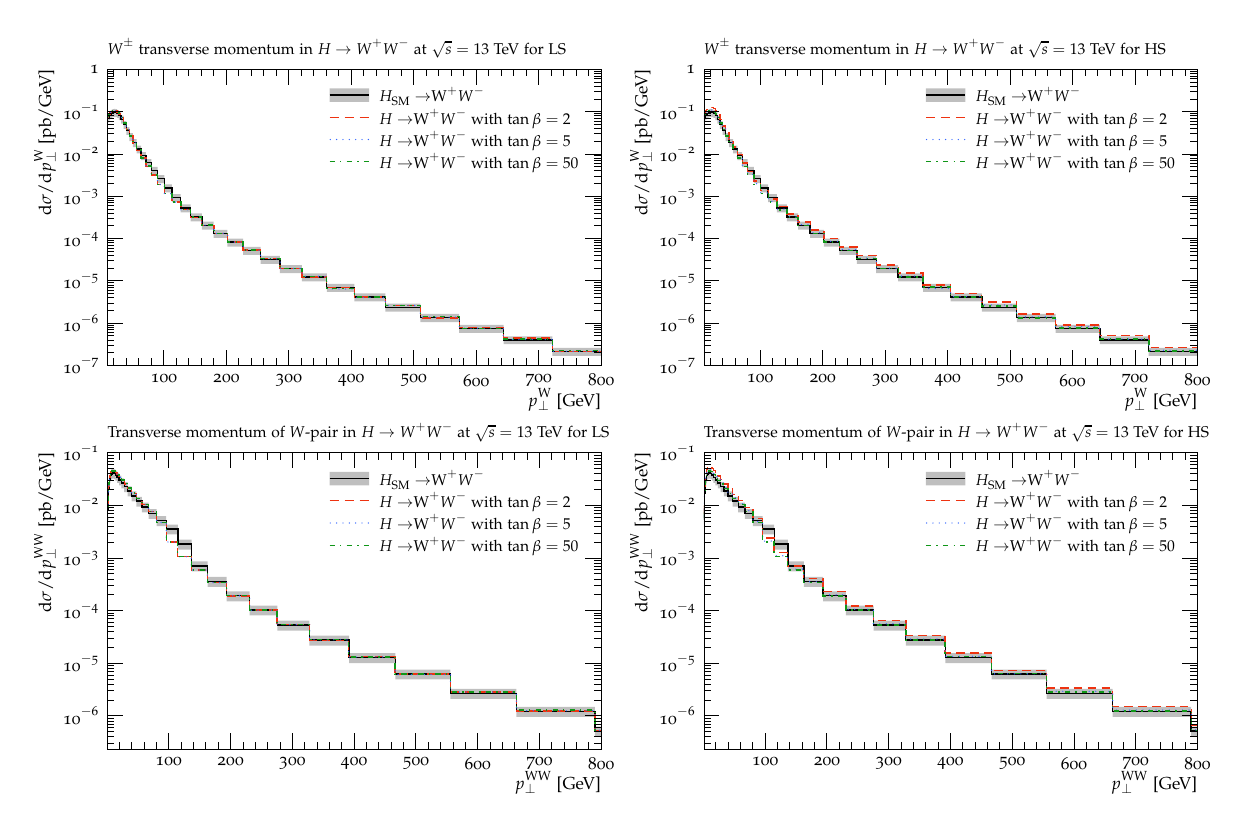}
\caption{\it The same as in Figure~\ref{WW-y}, but for differential cross-section as a function of transverse momentum for $W^{\pm}$-pair productions.}
\label{WW-pt}
\end{figure}

\begin{figure}[t]
\centering
\includegraphics[width=1\textwidth]{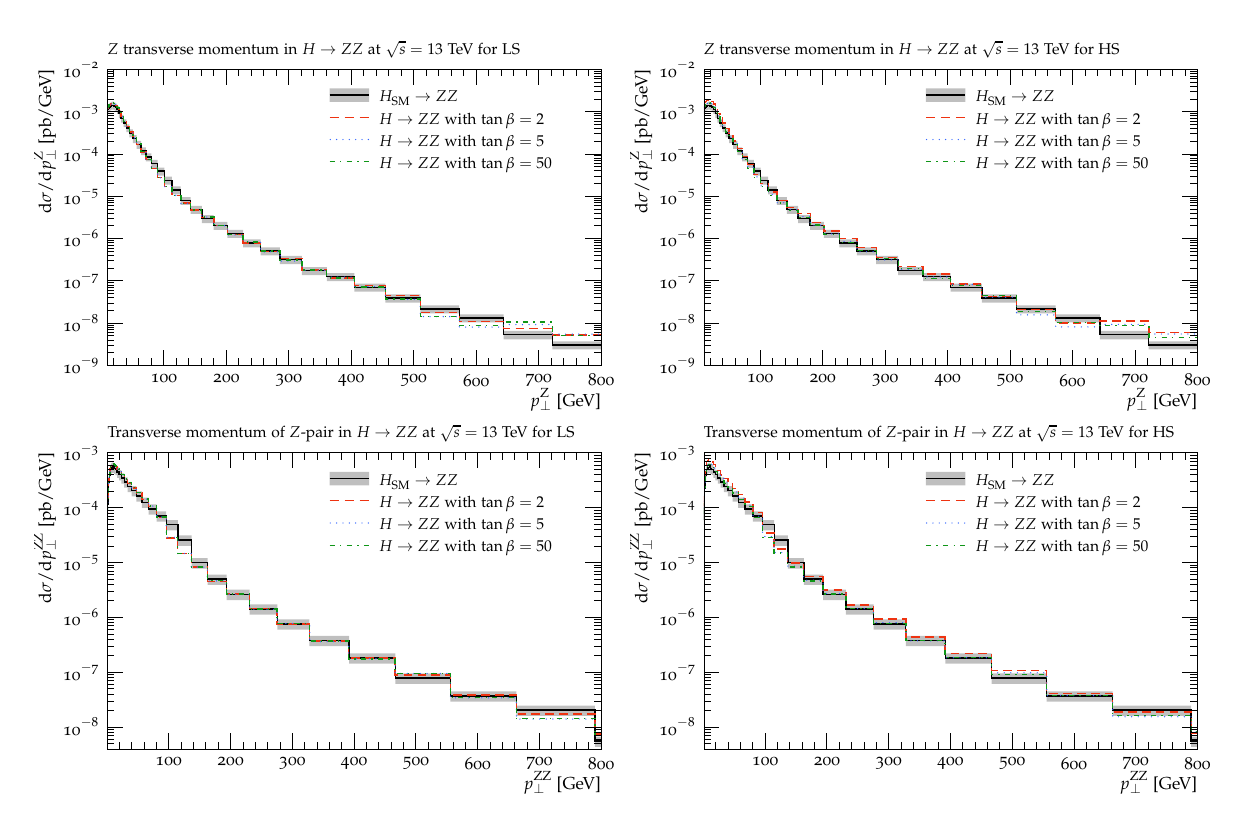}
\caption{\it The same as in Figure~\ref{WW-y}, but for differential cross-section as a function of transverse momentum for $Z$-pair productions.}
\label{ZZ-pt}
\end{figure}

In the next step, we calculate the single and the double Higgs production events within the MS-2HDM. According to our discussion in Section \ref{sec:alignment}, the MS-2HDM has two conformally-invariant quartic coupling unification points~$\mu^{(1)}_X$ (low-scale) and $\mu^{(2)}_X$ (high-scale), for a given choice of the charged Higgs-boson mass $M_{h^\pm}$ and $\tan\beta$~\cite{Darvishi:2019ltl,Darvishi:2020teg}. Thus, we perform these analysis for both unification points. In Figures~\ref{WW-y} and ~\ref{ZZ-y}, we display the differential cross-sections for $pp \to H \to VV^*$ production as a function of pseudo-rapidity of the produced gauge bosons in the MS-2HDM. The results for the low-scale (LS) and the high-scale (HS) points are shown in the left and the right panels, respectively. The top panels correspond to the kinematic properties of the EW bosons while the bottom panels depict the behaviour of gauge boson pairs. The calculations have been done at $13$~TeV center-of-mass energy for $M_{h^\pm}=500~\rm GeV$ and the typical values of $\tan\beta$, such as $\tan\beta=2,\,5$ and $50$, relevant to the benchmarks of Table~\ref{ex}. By analogy, Figures~\ref{WW-pt} and \ref{ZZ-pt} show the transverse momentum distributions for $W^{\pm}$ and $Z$ bosons production through single SM-like Higgs bosons. 

As expected, we observe that the results of $pp \to H \to W^+W^-$ and $pp \to H \to ZZ$ for both lower- and higher-scale quartic coupling unification points are in excellent agreement with the SM and the experimental data. Obviously, this is since the normalised couplings $g_{HVV}$ approaches the SM value $g_{H_{\text{SM}}V V} =1$ for both points. However, the deviation of the MS-2HDM results from the SM predictions is higher because the degree of misalignment reaches its maximum value for $\tan \beta=2$, while still remaining within their $1\sigma$ uncertainty. 

\begin{figure}[t]
\centering
\includegraphics[width=1\textwidth]{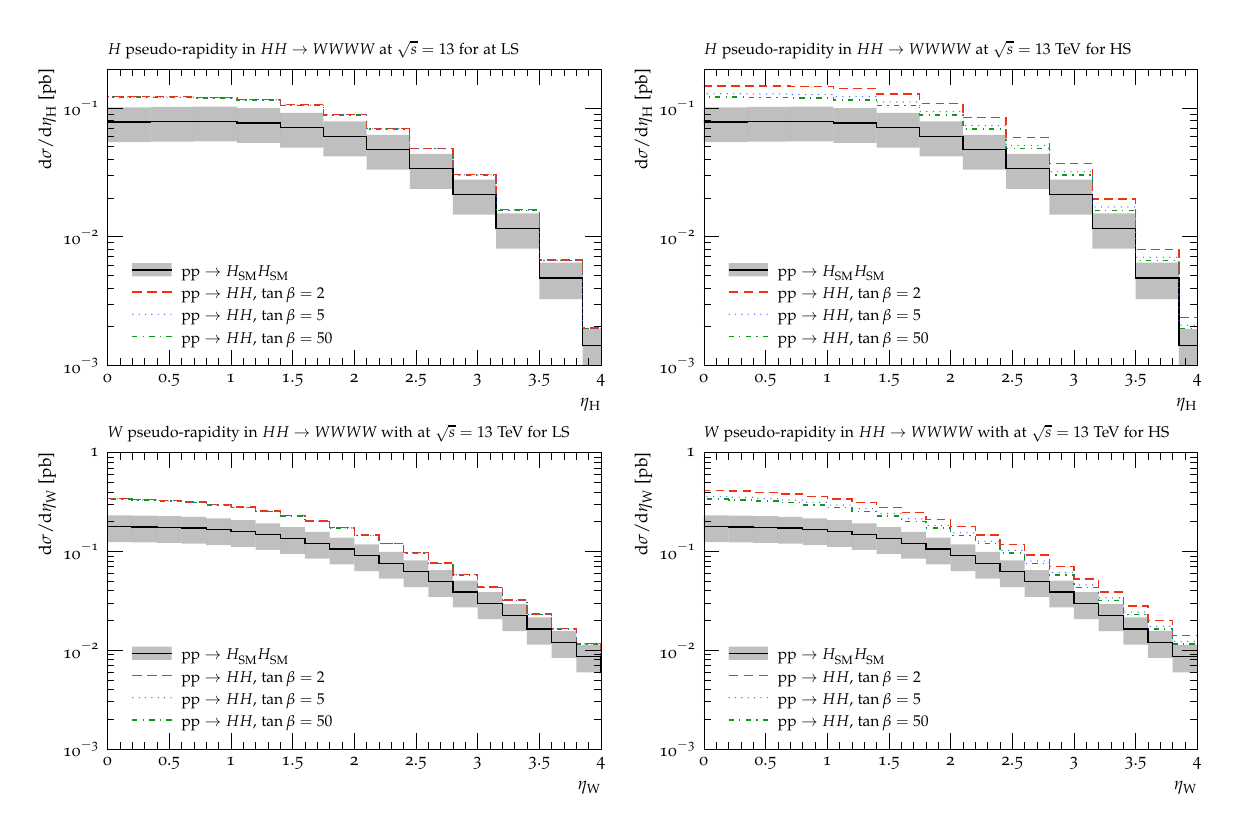}
\caption{\it Differential cross-section as a function of pseudo-rapidity for $W^{\pm}$-quadruplet productions through double SM-like Higgs boson ($H$) events  displayed for lower-scale  (higher-scale) quartic coupling unification point in left panels (right panels).  These are shown for  different values of $\tan \beta$ at $\sqrt{s}=13$~TeV. The top panels correspond to the kinematic properties of the $W^{\pm}$ bosons while the bottom panels depict the behaviour of $W^{\pm}$-pairs. Note that, these results are compared with relevant theoretical predictions in the SM within its uncertainty bounds.}
\label{HH_WW_y}
\end{figure}

\begin{figure}[t]
\centering
\includegraphics[width=1\textwidth]{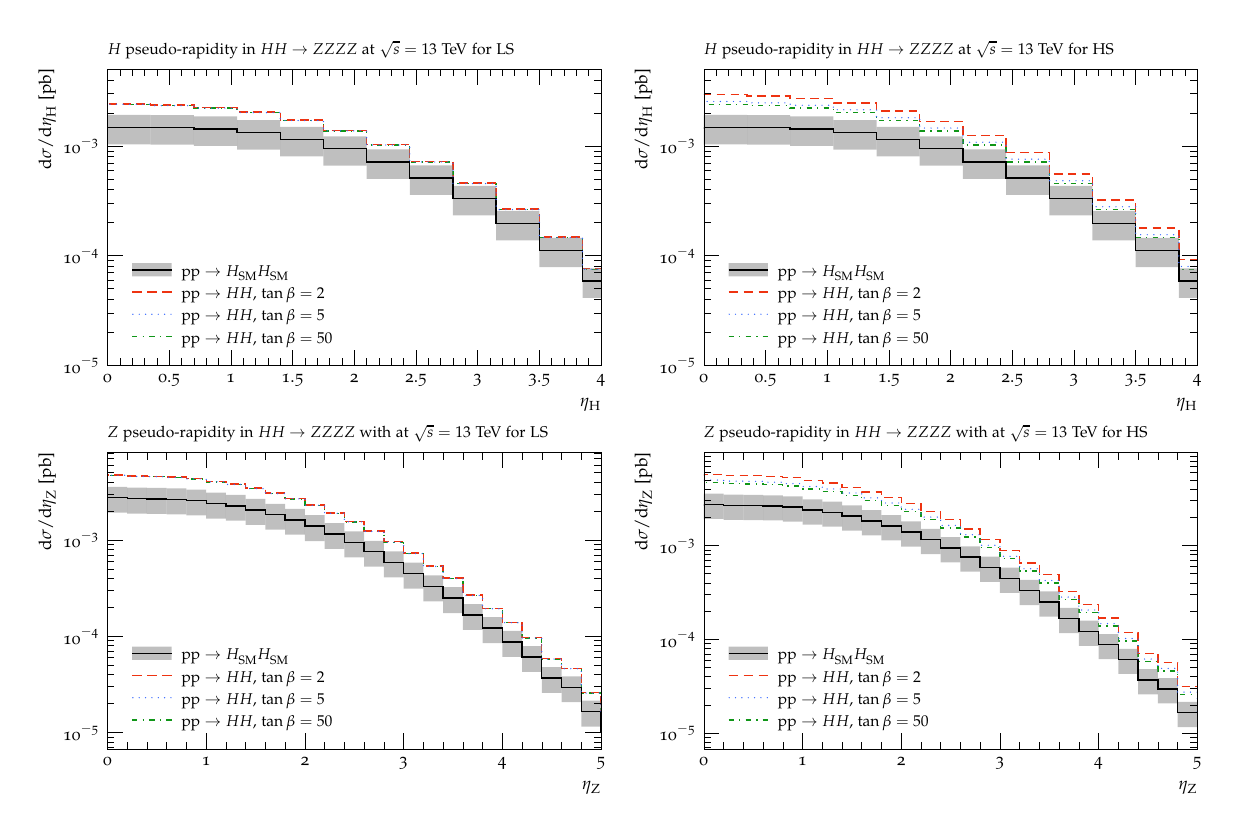}
\caption{\it The same as in Figure~\ref{HH_WW_y}, but for  $Z$-quadruplet productions.}
\label{HH_ZZ_y}
\end{figure}

Now, let us turn our attention to the $W^\pm/Z$-quadruplet production events through double Higgs decay channels, i.e. $pp \to HH \to VV^*V'V'^*$. Here, the sub-processes involving $H/h\to H H$ decays may have large contributions into the $p p \to H H$ production rate, as shown in Figure~\ref{fig1}(c).  
In Figures~\ref{HH_WW_y} and \ref{HH_ZZ_y}, we exhibit our results for the production of these events with $\tan \beta=2$, $5$ and $50$ at $13$ TeV center-of-mass energy. In both figures, the top panels represent the reconstructed kinematics of the exchanged $H$ scalars while bottom plots show the rates of production as functions of the pseudo-rapidity of the gauge bosons. These are shown for both the low-scale and the high-scale unification points. From these plots, we observe a substantial increase in the production rate of the SM-like Higgs bosons compared to their SM counterparts. This becomes more pronounced for the smaller values of $\tan \beta$, which has a nearly 2-fold increase compared to similar SM prediction. Despite the fact that these processes have expectedly smaller cross-sections in comparison with the single Higgs production events, they have substantial deviance from the SM and may be directly observed at the LHC data.

\begin{figure}[t]
\centering
\includegraphics[width=1\textwidth]{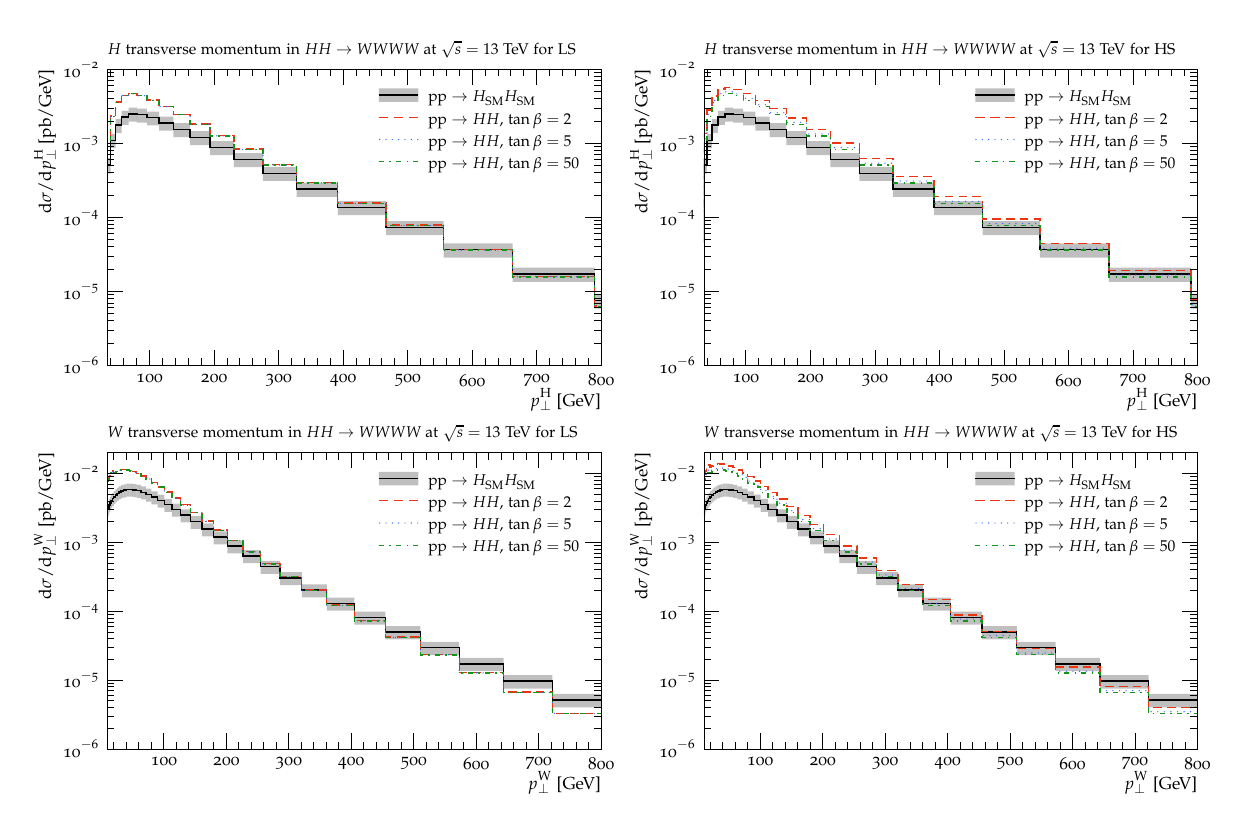}
\caption{\it The same as in Figure~\ref{HH_WW_y}, but for differential cross-section as a function of transverse momentum for $W^{\pm}$-quadruplet productions.}
\label{HH_WW_pt}
\end{figure}

\begin{figure}[t]
\centering
\includegraphics[width=1\textwidth]{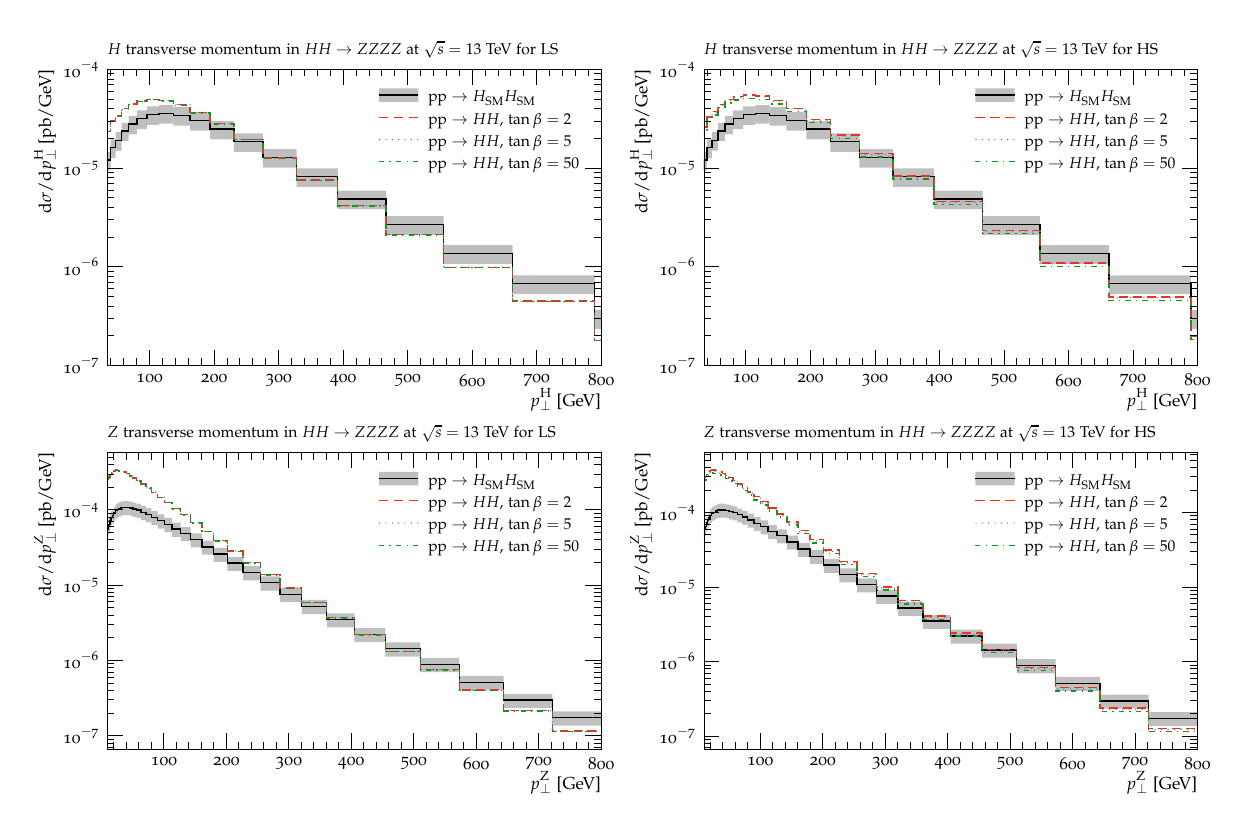}
\caption{\it The same as in Figure~\ref{HH_WW_y}, but for differential cross-section as a function of transverse momentum for $Z$-quadruplet productions.}
\label{HH_ZZ_pt}
\end{figure}

In a similar fashion, Figures~\ref{HH_WW_pt} and \ref{HH_ZZ_pt} demonstrate the transverse momentum distributions of the differential cross-section for the double Higgs production events. Observe that, the peaks in the Higgs and $W^\pm/Z$ bosons $p_{\perp}$ distributions are increased by a factor $\sim$3  with respect to the SM for $p_{\perp}<200$~GeV. However, the $p_{\perp}$ distribution's tails converge to the SM predictions in the high-$p_{\perp}$ regions. Therefore,  the signature of the MS-2HDM may be observed in the low-$p_{\perp}$ regions of the $W^\pm/Z$-quadruplet production events through double Higgs decay channels. Our observations can be readily generalised to other realisations of the 2HDM in their alignment limits.
 
\begin{figure}[t]
\centering
\includegraphics[width=1\textwidth]{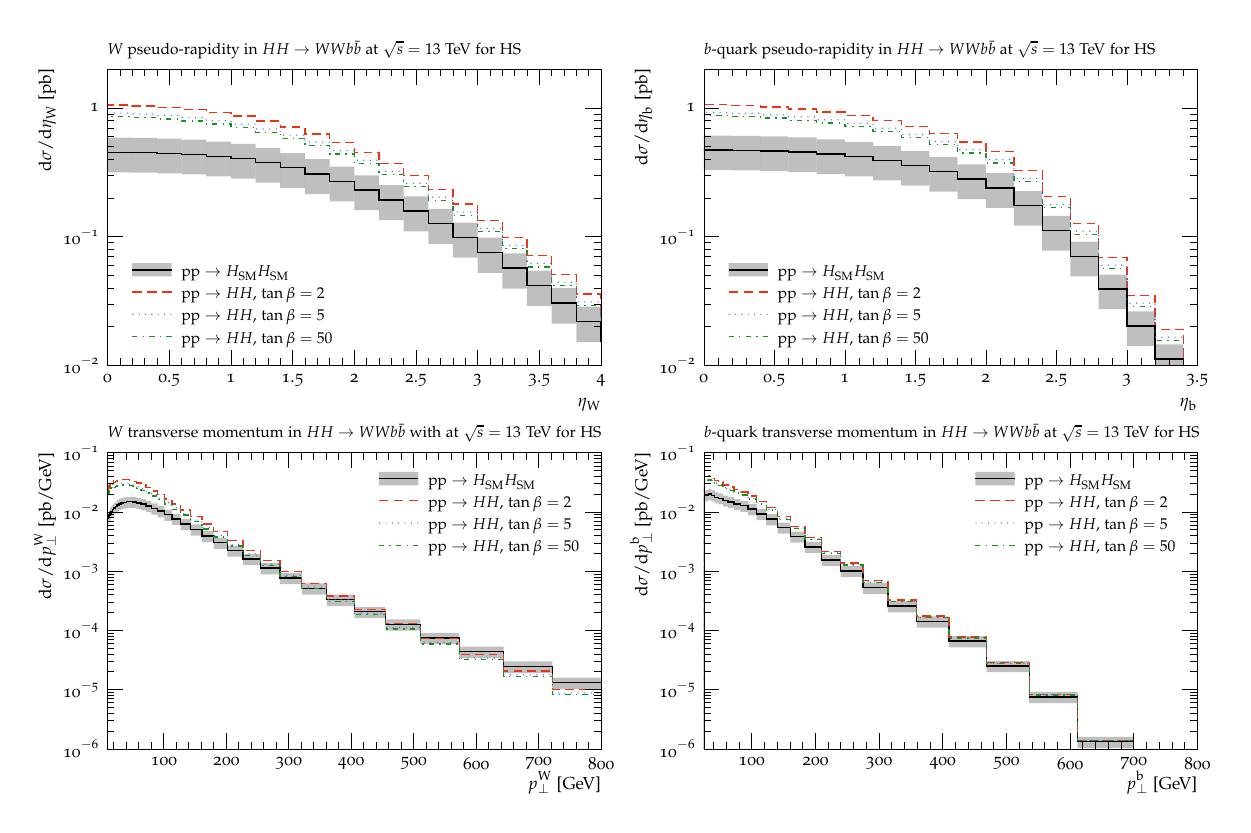}
\caption{\it Differential cross-section for the production of Higgs pairs with the subsequent $HH \to W^+W^-b\bar{b}$ decays. The differential cross-sections are plotted as functions of pseudo-rapidity (top panels) and transverse momentum (bottom panels) of the $W^{\pm}$ bosons (left panels) and $b$-quarks (right panels) productions for higher-scale quartic coupling unification point. These are shown for different values of $\tan \beta$ at $\sqrt{s}=13$~TeV. Note that, these results are compared with relevant theoretical predictions in the SM within its uncertainty bounds.}
\label{WWBB}
\end{figure}

As a final remark, it is also possible to look for the signature of the MS-2HDM through other significant signals like $b \bar{b} b \bar{b}$ and $W^+W^-b \bar{b}$ decay channel. In Figure~\ref{WWBB}, we show the differential cross-section for the production of Higgs pairs with the subsequent $HH \to W^+W^-b\bar{b}$ decays. It is noticeable that the resulting signals are expectedly larger compared to their counterparts in the $HH \to VV^*V'V'^*$ channels. This can be contributed to the larger branching ratio of $Br(H\to b\bar{b}) \sim \%57$ compared to $Br(H \to W^+W^-/ZZ) \sim\%$25 as well as direct contributions of the heavy neutral Higgs bosons through $h\to b\bar{b}$ ($a\to b\bar{b}$) decay mode with $Br(h \to b \bar{b}) \sim\%$5 ($Br(a \to b \bar{b}) \sim\%$2). However, these signals despite producing numerically larger contributions are plagued with very large SM backgrounds. In contrast, the $VV^*V'V'^*$ decay channels with their leptonic final states produce large and clean signatures for the MS-2HDM that might be observed in the out-coming LHC data.

\section{Conclusions}
\label{sec:Conclusions}

The data collected from CERN's LHC impose constraints over the coupling strengths of the Higgs boson, primarily to the EW gauge bosons ($V=W^{\pm},Z$), which are very close to the SM predictions. This simple fact severely restricts the form of possible scalar-sector extensions of the SM. In this study, we have considered the MS-2HDM where the SM alignment can be achieved naturally by the virtue of an SO(5) symmetry imposed on the 2HDM. 
The MS-2HDM is a minimal and very predictive extension of the SM governed by only three parameters: the unification scale~$\mu_X$, the charged Higgs mass $M_{h^{\pm}}$ and $\tan\beta$ which allow one to determine the entire Higgs sector of the model.

Given the remarkable features of the MS-2HDM~\cite{Darvishi:2019ltl,Darvishi:2020teg}, we have investigated the possible signature of this model via $W^{\pm}/Z$-quadruplet productions at the LHC. We have performed our calculations with NLO QCD accuracy for $p p \to HX \to VV^* X$ and $p p \to HHX \to VV^*V'V'^* X$ processes  for different values of $\tan \beta$, using the \textsf{Herwig~7} multi-purpose event generator at $\sqrt{s}=13$~TeV center-of-mass energy. 
The corresponding amplitudes are provided by \textsf{MadGraph5}, up to one QCD loop and two jets. The produced underlying events are showered by an AO \textit{MC@NLO} matched \textit{QCD+QED+EW} parton shower and the results have been analysed using \textsf{Rivet}.

We have shown that the predictions for $W^{\pm}/Z$-pair productions through single SM-like Higgs boson events are aligned with their SM counterparts, while the presence of the heavy Higgs states significantly enhances the $W^{\pm}/Z$-quadruple productions.  Particularly,  we have found that the cross-section for these events is increased by a factor $\sim$3 in $p_{\perp}<200$~GeV region with respect to the SM. However, these distributions converge to the SM predictions in the high-$p_{\perp}$ regions. Therefore,  the signature of the MS-2HDM may be observed in the low-$p_{\perp}$ regions of the $W^\pm/Z$-quadruplet production events through double Higgs decay channels. These observations are very helpful toward the possible future discovery of this model.

\begin{acknowledgments}
\noindent

The authors would like to thank Prof. A.~Pilaftsis and Prof. P.~Richardson for their instructive discussions.
\textit{ND} is supported by the Lancaster-Manchester-Sheffield Consortium for Fundamental Physics, under STFC research grant ST/P000800/1. 
The work of \textit{ND} is also supported in part by the National Science Centre 
(Poland) as a research project, decision no 2017/25/B/ST2/00191 and by the Polish National Science Centre HARMONIA grant under contract UMO-2015/20/M/ST2/00518 (2016-2020).
\textit{MRM} is supported by the UK Science and Technology Facilities Council (grant numbers ST/P001246/1). 
This work has received funding from the European Union's Horizon 2020 research and innovation program as part of the Marie Sk\l{}odowska-Curie Innovative Training Network MCnetITN3 (grant agreement no. 722104).

\end{acknowledgments}

\end{document}